\pdfoutput=1

\documentclass[apjl]{emulateapj}

\usepackage{color}

\shorttitle{Plasma transport across the heliopause}
\shortauthors{Strumik et al.}

\begin{document}


\title{ADVECTIVE TRANSPORT OF INTERSTELLAR PLASMA INTO THE HELIOSPHERE ACROSS THE RECONNECTING HELIOPAUSE}


\author{M. Strumik$^1$, S. Grzedzielski$^1$, A. Czechowski$^1$,
W. M. Macek$^{1,2}$ and R. Ratkiewicz$^{1,3}$}
\affil{$^1$Space Research Centre, Polish Academy of Sciences, Bartycka 18A, 00-716 Warsaw, Poland \\
$^2$Faculty of Mathematics and Natural Sciences. Cardinal Stefan Wyszy\'{n}ski University, W\'{o}ycickiego 1/3, 01-938 Warsaw, Poland \\
$^3$Institute of Aviation, Al. Krakowska 110/114, 02-256 Warsaw, Poland}




\begin{abstract}
We discuss results of magnetohydrodynamical model simulations of plasma dynamics in the
proximity of the heliopause (HP). The model is shown to fit details
of the magnetic field variations observed by {\it Voyager 1} spacecraft during the transition
from the heliosphere to the local interstellar medium (LISM). We
propose an interpretation of magnetic field structures observed by
Voyager 1 in terms of fine-scale physical processes. Our simulations
reveal an effective transport mechanism of relatively dense LISM plasma
across the reconnecting HP into the heliosphere. The mechanism is associated
with annihilation of magnetic sectors in the heliospheric plasma near
the HP.
\end{abstract}





\section{Introduction}

The relative motion of the Sun with respect to the local
interstellar medium (LISM) leads to the formation of a cavity in the
ambient interstellar medium called the heliosphere. The solar wind
(SW) and the LISM plasma flow are assumed to be separated by the
heliopause (HP), located between a termination shock (TS) in the
SW and possibly a bow shock (BS) in the LISM. The inner
heliosheath (IHS) is defined to be located between the TS and HP, while
the outer heliosheath (OHS, located between the HP and BS, if the BS
exists) is a region, where significant modification of the
LISM flow occurs.

Recent measurements of the {\it Voyager 1}
({\it V1} hereafter) spacecraft  provided puzzling observational data
that resulted in controversy concerning their interpretation. The
{\it V1} spacecraft observed two partial depletions in anomalous
cosmic ray (ACR) fluxes, followed by a decrease to the
instrumental background at $\sim$ 122 AU from the Sun \citep{Krietal13}.
The variations in the ACRs were anticorrelated with changes in the galactic
cosmic ray (GCR) fluxes that significantly increased
\citep{WebMcD13,Krietal13,Stoetal13}. At the same time sudden
enhancements in the magnetic pressure were observed,
but the lack of significant change in the direction of the
magnetic field vector has led to doubts whether the
observed changes were associated with the HP \citep{Buretal13}. However,
theoretical studies of the magnetic reconnection at the HP
based on numerical simulations have suggested that the {\it  V1} observations are
consistent with the crossing of a structure related to the HP modified
by magnetic reconnection \citep{Swietal13,Stretal13}.
These doubts were finally dispelled with the detection of local plasma oscillations
by the {\it V1} plasma wave instrument, where the deduced value of
the number density of the surrounding plasma clearly indicated that {\it V1} has
entered the interstellar medium \citep{Guretal13}.
In view of the recent {\it V1} observations the
question of the structure of the HP and fine-scale phenomena
around the discontinuity surface has become an interesting and 
timely problem.

We present results of numerical simulations that aim to give a detailed explanation of
the {\it V1} measurements of the magnetic field vector changes associated
with the ACR boundary crossing. Starting our simulation from a
simple configuration of two laminar current sheets (one of them representing the HP) we calculate the time
evolution of the plasma parameters and the magnetic field in
an area of linear size 4 AU in the normal direction to the HP.
Since the linear scale of our
computational problem is large in comparison with the ion inertial
length $\lambda_\mathrm{i}=V_\mathrm{A}/\omega_\mathrm{ci} \lesssim
10^{-5}$ AU, we use a magnetohydrodynamical (MHD) approach.
Length scales considered in our Letter are large in comparison with the
tens or hundreds of $\lambda_\mathrm{i}$ typical for kinetic (particle-in-cell, PIC)
simulations of the magnetic reconnection in the heliosphere \citep{Draetal10,Swietal10,Swietal13}.
The PIC simulation results can be linked to larger scales by scaling arguments \citep{Schetal12,Swietal13},
but a fully consistent description of processes on larger (several AU) scales is feasible at present by using the MHD approach.
Note also that the scales considered in our Letter are
much smaller than the characteristic scales of the HP instabilities caused
by charge exchange \citep{Lieetal96,Zanetal96,Floetal05,Boretal08}.

It has been suggested that (for the LISM magnetic field of the order of a fraction of nT)
reconnection processes at the HP may play an important role
in mixing between heliospheric and interstellar plasmas \citep{MacGrz85,Fahetal86}.
We discuss in this context properties of advective transport of the LISM plasma into
the heliosphere that is revealed by our simulations.

\section{Model}

MHD equations are solved numerically in 2.5 dimensional geometry
(vanishing out-of-plane derivatives) using the high-resolution MUSCL
scheme \citep{KurTad00}. Resistive and viscous effects are not
included explicitly but result from small numerical diffusion of the
numerical scheme. A divergence-free magnetic field is ensured by applying
a flux-constrained (staggered mesh) approach \citep{BalSpi99}. Since
phenomena related to the interaction of plasma and neutrals are typically
associated with much larger spatial scales, no neutral particle
background is included in our model. Periodic boundary
conditions are applied in the $x$-direction and open boundary conditions
in the $y$-direction.
Velocity is normalized to the Alfv\'en speed $V_\mathrm{A,IHS}$ in the IHS.
The MHD equations are scale-less, which allows us to choose the length unit
to be 1 AU, the resulting time unit is then $T_0=$AU/$V_\mathrm{A,IHS} \approx 17.5$ days.
The resolution of the simulation grid is
3840$\times$768 points and the computational domain size is $L_x
\times L_y = 20~\mathrm{AU} \times 4~\mathrm{AU}$. The density and magnetic field in
the simulation are normalized to their averages in the
IHS, $N_0$ and $B_0$.

\begin{figure}[!htbp]
\begin{center}
\includegraphics[width=8.7cm]{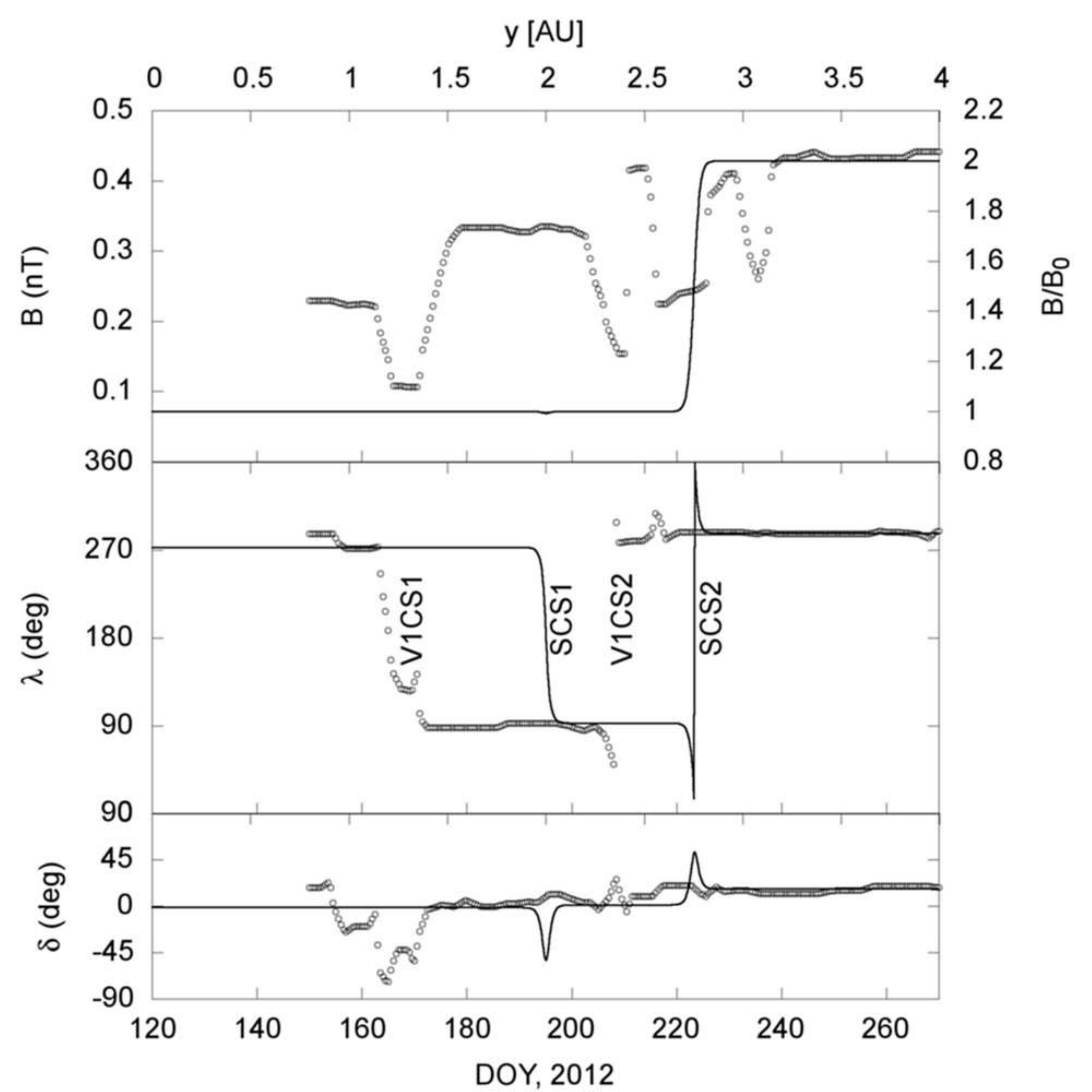}
\caption{Profiles of the magnetic field strength $B$, and the angles
$\lambda$ and $\delta$ measured by the {\it V1} spacecraft
(circles, left vertical and bottom horizontal axes). The solid line shows the profiles
set up using the initial condition along the $y$-axis 
in our simulation (right vertical and top horizontal axes).}
\label{profiles}
\end{center}
\end{figure}

\begin{figure}[!htbp]
\begin{center}
\includegraphics[width=8.5cm]{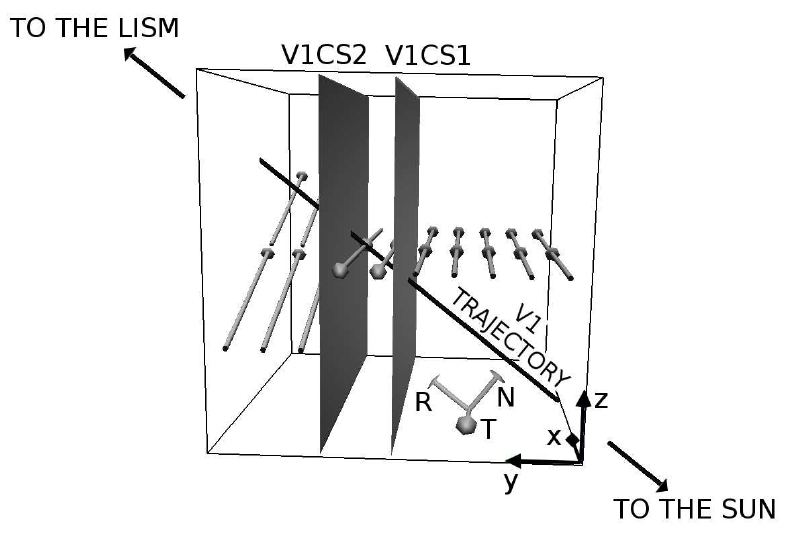}
\caption{
Orientation of the simulation frame $xyz$ with respect
to the $RTN$ frame, the current sheets V1CS1 and V1CS2 (normal
to the $y$-axis), and the {\it V1} trajectory (along the $R$-axis
approximately). The direction of the magnetic field lines is
shown by non-labeled gray arrows inside the box, the vector direction changes at the
current sheets V1CS1 and V1CS2.}
\label{xyz_rtn}
\end{center}
\end{figure}

The measurements of the magnetic field vector variations by the {\it V1}
spacecraft are conventionally presented in the $RTN$ frame (see, e.g.
\cite{BurNes12} and references therein). The orientation of the magnetic
field vector is given by two angles. The $\lambda$ angle describes
deviation of the magnetic field vector from the radial ($R$) direction in
the radial--tangential ($R$--$T$) plane, while the $\delta$ angle specifies the
deviation of the magnetic field vector from the $R$--$T$ plane. 
Figure \ref{profiles} shows the {\it V1} measurements of the
magnetic field vector variations
based on data that we digitally extracted from Figure 2 in
\citet{Buretal13}. In the extraction process the data are obtained
on a daily timescale, neglecting the details on the fine (48-s) timescale. In
Figure \ref{profiles} two crossings of current sheets
observed by {\it V1} are indicated by the $\lambda$ angle changing
from $\sim$270$^\circ$ to $\sim$90$^\circ$ (days 162-170 of 2012,
V1CS1 hereafter) and then from $\sim$90$^\circ$
to $\sim$270$^\circ$ (day 208 of 2012, V1CS2) and are associated
with variations of the $\delta$ angle and the
magnetic field strength $B$. Note that for the time interval shown in
Figure \ref{profiles}, {\it V1} was $\sim34^\circ$ north of the solar equator and
the magnetic polarity of the northern hemisphere of the
Sun was negative ($\lambda \approx 90^\circ$).
Minimum variance analysis (see,
e.g. \cite{SonSch98} for description of the method) suggests
that the normal vector to the current sheet V1CS2 has the direction
$(0.7923,0.0551,-0.6076)$ in the $RTN$ frame, which is close
to the direction of the unperturbed LISM flow. To include this particular
physical configuration in our simulations, we assume that the simulation
frame $xyz$ is related
to the $RTN$ frame through the following transformation
\begin{equation}
\left(
\begin{array}{c}
\hat{\mathbf{e}}_R \\
\hat{\mathbf{e}}_T \\
\hat{\mathbf{e}}_N
\end{array}
\right) = \left(
\begin{array}{rrr}
  0.0527 & 0.7923 & 0.6079 \\
  -0.9983 & 0.0551 & 0.0147 \\
  -0.0218 & -0.6076 & 0.7939
\end{array}
\right)
\left(
\begin{array}{c}
\hat{\mathbf{e}}_x \\
\hat{\mathbf{e}}_y \\
\hat{\mathbf{e}}_z
\end{array}
\right)
\end{equation}
Figure \ref{xyz_rtn} shows the relative orientation of the two frames with
respect to the current sheets and magnetic field lines arranged in the
initial condition of our simulation. Note that the transformation matrix
implies $x \approx -T$, $y$-axis is the normal vector to the current
sheets, and the angle between $y$- and $R$-axes
is $\sim$37$^\circ$. One should also note that the 2.5 dimensional geometry
assumption means that the $x$--$y$ plane is the simulation
domain and $\partial / \partial z = 0$ but the magnetic field and velocity
vectors may have three non-zero components in general.

Length scales for our simulations are roughly of the same order of magnitude
as the separation of the current sheets V1CS1 and V1CS2 observed by {\it V1}. The
{\it V1} speed ($\sim$17 km s$^{-1}$) and  the time interval ($\sim$ 43 days)
between the current sheets observations yield a separation $\sim$0.42 AU
with a large uncertainty resulting from neglecting an unknown mean plasma
flow. As illustrated in Figure \ref{profiles} (solid line), in the
initial condition for our simulations we set up two current sheets
SCS1 (at $y=2$ AU) and SCS2 (at $y=2.75$ AU), corresponding to the
current sheets V1CS1 and V1CS2 observed by the {\it V1} spacecraft.
We assume that the normal component
of the magnetic field $B_y$ is initially zero everywhere, and that
$B_x/B_0=1$ and $B_z/B_0=0$ for $y<2$ AU.
Appropriate rotations of the magnetic field vector in the $B_x$--$B_z$ plane
(by 180$^\circ$ at the SCS1 and 157$^\circ$ at the SCS2) are applied
to obtain the variations of the angles $\lambda$ and $\delta$ shown by
the solid line in Figure \ref{profiles}.
We interpret the region $y<2.75$ to be the IHS and
the remaining region to be the OHS. Therefore the current sheet SCS2 can be
associated with an initially laminar HP and SCS1 with an occurrence
of the heliospheric current sheet (HCS). For  SCS2 we arrange the magnetic
strength jump $B_\mathrm{OHS}/B_\mathrm{IHS}=2$ and the number density
jump $N_\mathrm{OHS}/N_\mathrm{IHS}=20$, similar to our previous
work \citep{Stretal13}. The magnetic strength jump is consistent with the
{\it V1} observations \citep{Buretal13}.
The magnitude of the number density jump is chosen to agree
with recent measurements in the IHS by {\it V2} that give $N_\mathrm{IHS}=0.0025$ cm$^{-3}$ \citep{RicWan12}
and with the OHS value $N_\mathrm{OHS}=0.05$ cm$^{-3}$ determined from the outset
of the upward drifting radio emission at frequency $\sim$1.8 kHz
as reported by \citet{Guretal13}. To ensure total (thermal+magnetic)
pressure equilibrium between the IHS and OHS regions we use $\Delta
\beta=\beta_\mathrm{IHS}-\beta_\mathrm{OHS}=3.3$ at SCS2
where $\beta=2 \mu_0 p/B^2$ is the ratio of the kinetic and magnetic
energy densities.
No mean flow in the simulation box is imposed in the initial condition.

To initiate magnetic reconnection we locally
impose small-amplitude random perturbations on the plasma pressure in
the proximity of the
SCS2 location in the simulation box. A locally increased level
of noise is supposed to accelerate the growth of the tearing instability
and the development of the reconnection sites.

\section{Simulation results and model validation}

The simulation starting from the initial condition described above results in
a time evolution, where the development of the reconnection sites
at random locations on SCS2 initiates the growth of magnetic
islands at the current sheet (see, e.g. \citet{Stretal13} for a more
detailed discussion of a qualitatively similar scenario).
Merging interacting magnetic islands is observed,
leading to the growth of the islands' size, while the number
of magnetic islands decreases in time. Fluctuations associated
with the reconnection at SCS2 propagate throughout the
simulation box and initiate reconnection at the current sheet SCS1.
At a later time, magnetic islands at the  two current
sheets start to interact. These dynamical processes cause plasma density
and magnetic field compressions and variations in the magnetic field and
velocity vectors.

\begin{figure}[!htbp]
\begin{center}
\includegraphics[width=8.5cm]{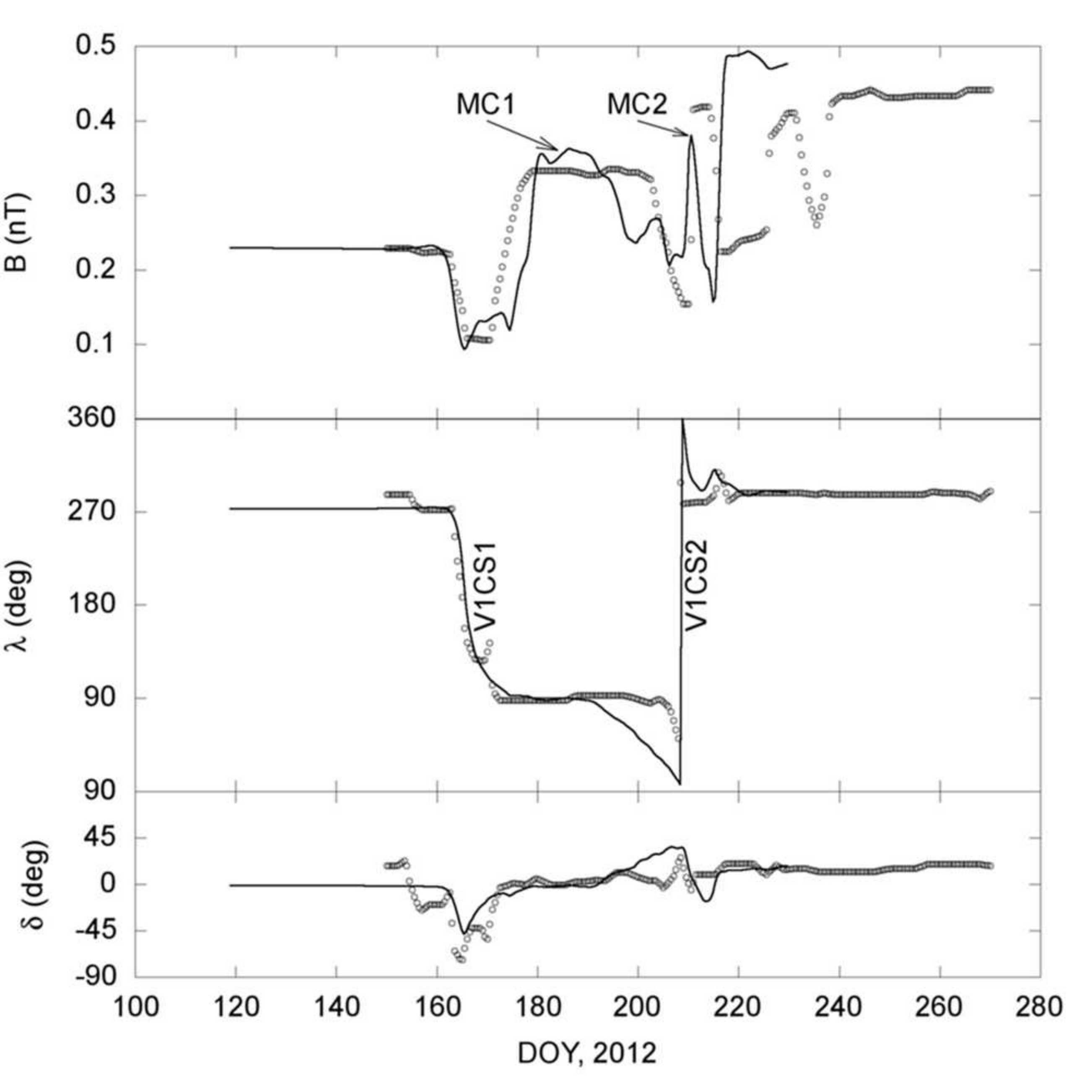}
\caption{Comparison of the profiles of the magnetic field strength
$B$, and the $\lambda$ and $\delta$ angles measured by the {\it V1} spacecraft
(circles) with corresponding profiles (properly shifted and rescaled)
obtained for a virtual spacecraft moving in the simulation box (solid line).}
\label{profiles_sim_data}
\end{center}
\end{figure}

Figure \ref{profiles_sim_data} shows comparison of the {\it V1} observations and
similar
profiles found in our simulations along a virtual spacecraft trajectory
$\mathbf{r}=\mathbf{r}_0+\mathbf{v}_\mathrm{SC}(t-t_0)$, where
in the simulation frame $xyz$ the initial
position is $\mathbf{r}_0=(5.35,1,0)$ AU, the virtual spacecraft velocity vector is
$\mathbf{v}_\mathrm{SC}/V_\mathrm{A,IHS}=(0.016,0.161,0.058)$, and $t_0/T_0=14.5$.
The virtual spacecraft speed $v_\mathrm{SC}/V_\mathrm{A,IHS} \approx 0.172$ in
the simulation corresponds to the {\it V1} speed $v_\mathrm{V1} \approx 17$ km s$^{-1}$.
The magnetic field vector is assumed constant in the $z$-direction
in the simulation and linear interpolation in the space domain (based on the
simulation grid) and time domain (using frames separated by $\Delta t=0.1/T_0$) is used
to obtain the magnetic field vector components between nodes of the simulation grid.
Note that the profiles presented in Figure \ref{profiles_sim_data} result from the motion
of the virtual spacecraft in a spatially inhomogeneous environment accompanied
by temporal changes of the spatial distribution of the magnetic field vector.
As illustrated in Figure \ref{profiles_sim_data} the simulation properly reproduces the
following features of the {\it V1} measurements:
the magnetic field enhancement
of a factor 1.4 between the V1CS1 and V1CS2 crossings (marked by MC1),
and strong pulsations of the magnetic field strength after the V1CS2
crossing (marked by MC2).
The V1CS1 crossing is associated
with $\delta$ angle pulsations in the negative range
of values, whereas mostly positive values of $\delta$ were measured for the V1CS2
crossing. One can see that the simulated $\delta$ profiles correspond to
the {\it V1} data.

\begin{figure}[!htbp]
\begin{center}
\includegraphics[width=8.5cm]{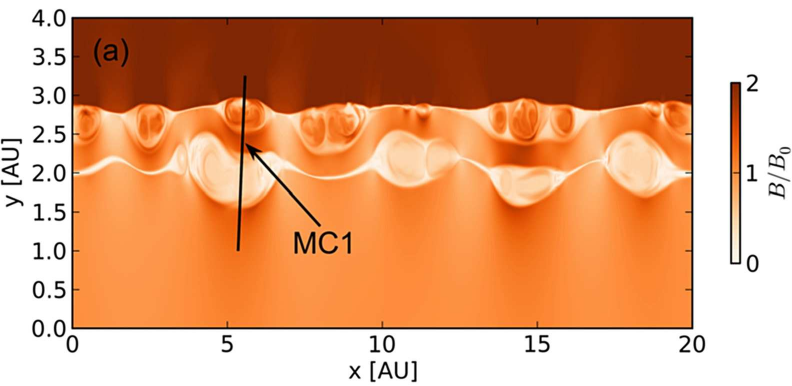}
\includegraphics[width=8.5cm]{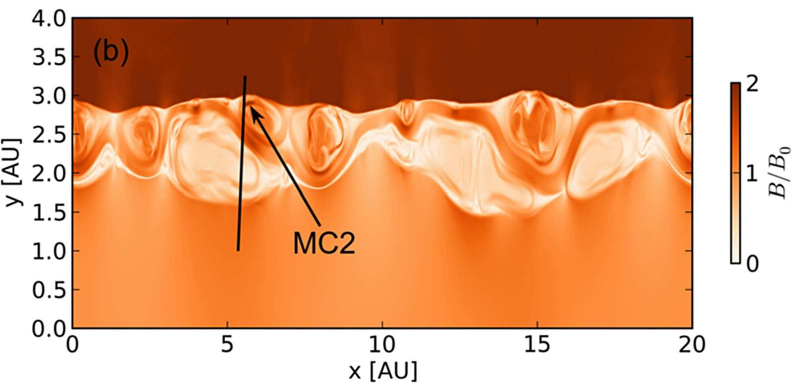}
\caption{Development of magnetic islands and magnetic compressions in the
simulation for (a) $t/T_0=23.0$ and (b) $t/T_0=26.1$. The magnetic islands
can be seen quite clearly in panel (a)
as local thickenings of the current sheets SCS1 ($y \approx 2$,
low values of $B/B_0$) and SCS2 ($y \approx 2.75$, sharp gradient of $B/B_0$).
The non-labeled black line shows the $x$--$y$ plane projection of the 
virtual spacecraft trajectory for which profiles from 
Figure \ref{profiles_sim_data} are obtained.
Locations of the
magnetic compressions MC1 and MC2 from Figure \ref{profiles_sim_data} are 
shown by arrows. The aspect
ratio of the plot has been altered to better visualize details.}
\label{f4}
\end{center}
\end{figure}

Further inspection of the numerical solution allows us to identify
the fine-scale physical effects responsible for the appearance
of the observational features discussed above.
Figure \ref{f4}(a) shows the spatial distribution of $B$ for $t/T_0=23$
(occurrence of MC1), and Figure \ref{f4}(b) shows the same but for $t/T_0=26.1$
(occurrence of MC2).
The $\sim$1.4 fold increase
of the magnetic field strength (MC1 in Figure \ref{profiles_sim_data}) between V1CS1 and V1CS2 is
apparently related to a magnetic compression occurring when two magnetic islands
from neighboring current sheets collide with each other, as
illustrated in Figure \ref{f4}(a).
This is
a rather typical feature in this physical configuration as resulting from
independent motion of magnetic islands at neighboring current
sheets during the early stage of development of the islands. When the
islands at two current sheets have grown to a size comparable to the
distance between the sheets, they start to
interact causing significant magnetic compressions. Even
stronger magnetic compressions are seen after the
V1CS2 crossing, marked by MC2 in Figure \ref{profiles_sim_data}.
As shown in Figure \ref{f4}(b) these magnetic strength
enhancements are associated with the internal structure of
the magnetic islands, appearing during the merging process
of two islands significantly different in size. It is again a rather
typical situation that reconnection sites initiated at
random locations on a current sheet produce magnetic islands that
have generally different sizes. Therefore the merging process is
likely to include magnetic islands of different sizes.

Due to the purely MHD nature of our simulations the cosmic ray component is not included
in our computations. Therefore accurate
tracking of high-energy particle fluxes is not possible.
Heuristically we may argue that since the magnetic compression MC2 in the simulation
is part of a magnetic island that has grown at SCS2 discontinuity,
it is likely to contain higher energy particles from the LISM and a
deficit of cosmic rays of heliospheric origin. However, a detailed
description of cosmic ray fluxes is beyond the scope of this Letter.


\section{Transport of plasma across the HP}
\label{sec:transport}
\begin{figure}[!htbp]
\begin{center}
\includegraphics[width=8.7cm]{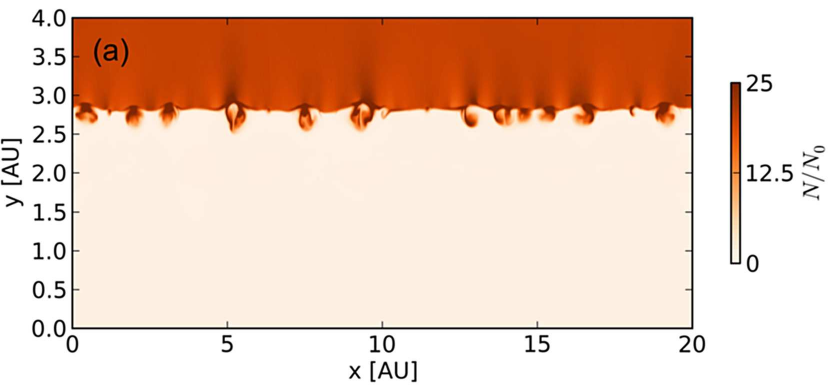}
\includegraphics[width=8.7cm]{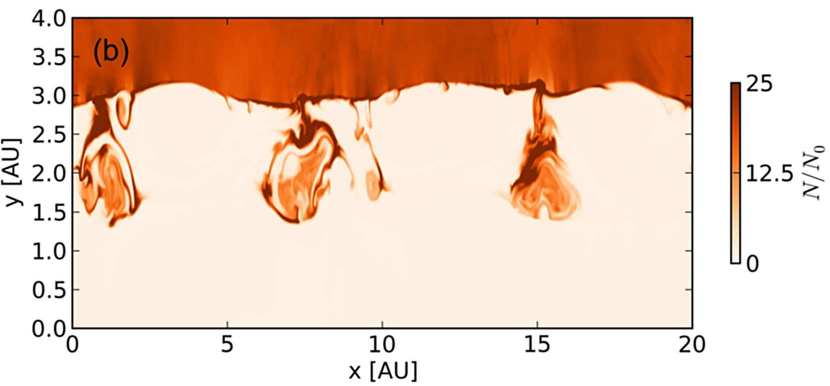}
\includegraphics[width=8.7cm]{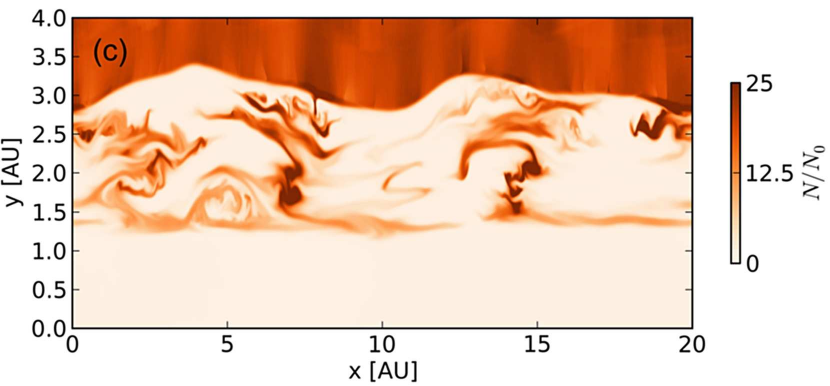}
\caption{Development of the density intrusions caused by the transport of the dense LISM plasma
to the IHS for (a) $t/T_0=20$, (b) $t/T_0=35$, and (c) $t/T_0=60$. The aspect
ratio of the plot has been altered to better visualize details.}
\label{f5}
\end{center}
\end{figure}

Another novel feature revealed by our simulation
is advective transport of the relatively dense LISM plasma
into the heliosphere through the reconnecting HP. Due to the magnetic strength
and density jump across the current sheet SCS2 we obtain a bulge of
magnetic separatrices and reconnection jets toward the IHS
(small values of $y$), which is a typical feature of
asymmetric reconnection (see, e.g., \citet{Swietal10}). During the
development of the reconnection sites in random locations on the current
sheet, reconnection jets from different reconnection sites
interact causing local density and pressure enhancements.
The plasma flow velocity component is $V_y<0$ on average in these compressions, and thus plasma
is transported away from the HP toward the IHS (see Figure \ref{f5}(a), $y<2.75$), which
is the lower density region. At a later moment in time, the transport of plasma across
the SCS2 occurs in several ducts (see Figure \ref{f5}(b)) spontaneously appearing
on the current sheet as a result of merging of smaller structures. When magnetic
islands from the two current
sheets in the simulation start to interact, plasma
from the density intrusions is transported effectively deep into the IHS.
Later, the magnetic sector initially set up between the SCS1 and SCS2 is
annihilated and the reconnection process and transport
of the dense plasma across the SCS2 ceases. As illustrated in Figure
\ref{f5}(c), in the final state we observe dense plasma intrusions
separated from the SCS2 and layered in the IHS region.
Our results indicate that the IHS penetration depth for the
intrusions is $\sim 1.5$ AU, but the simulation does not include the IHS
turbulence that could presumably transport the density intrusions even
deeper into the IHS. The average flux inside the transport ducts estimated
from our computations is $\sim7.5 \times 10^5$ cm$^{-2}$ s$^{-1}$,
the typical linear size of the transport ducts
is $\sim$0.5 AU, the separation distance between the ducts is $\sim$7 AU,
and the transport event duration time is $\sim$1.5 yr.

\section{Conclusions}

We present results of the modeling of the plasma dynamics in the proximity of
the HP that shed light on its internal structure and the surrounding region. The
model is validated by finding virtual spacecraft
observations corresponding to the magnetic field vector measurements
obtained during {\it V1} crossing through the ACR boundary. This
allows us to provide an interpretation of particular segments of the
observations in terms of the fine-scale physical effects responsible for
their appearance. Results of our simulations suggest that characteristic
features of the {\it V1} spacecraft measurements can be explained as
resulting from dynamical processes driven by magnetic reconnection
occurring at two closely separated current sheets, the HCS and the HP.
Our simulation also reveals a transport mechanism of the relatively
dense LISM plasma through the HP to the IHS region, that may
be responsible for the formation of density intrusions in the IHS filled by
dense plasma of interstellar origin. The values of the basic
parameters characterizing the transport process are reported in
Section \ref{sec:transport} of our Letter.

The advective transport of plasma has been obtained
within an MHD approach that describes the dynamics of the thermal plasma component.
The plasma experiment for {\it V1} has not been in operation since 1981, and
thus plasma density measurements are not available for 2012.
The density estimates deduced by \citet{Guretal13} from the plasma wave instrument 
were obtained for much later times than the final jump in the magnetic field strength
(seen on day 237 of 2012). The region of the density intrusions is predicted by our
computations to appear ahead of the jump, thus the measurements by \citet{Guretal13}
do not allow for observational confirmation of the transport process.
Mechanisms of transport of high-speed cosmic rays are generally different from 
the advective transport of thermal plasma described by the MHD approximation.
Therefore, it is not clear at present how the density intrusions revealed by our simulations
may be related to abrupt jumps of ACR or GCR fluxes observed by {\it V1} (days
210--237 of 2012). Possible HP crossing by {\it V2} in the future could provide
interesting data in this respect.

Our results are interesting in the context of the so-called pile-up problem for the
HCS. The HCS foldings are constantly produced by the changing magnetic
polarity of the Sun and advected by the SW to the IHS region
\citep{Neretal95,Czeetal10b}. The recent {\it V1} observations indicate that
at least some of the HCS foldings and the magnetic sectors between them may survive the advective transport in
turbulent SW and appear very close to the HP. This suggests that
the annihilation of magnetic sectors in the direct heliospheric
neighborhood of the HP could be a quite likely solution to the pile-up problem for
the HCS. For this reason one may expect that the scenario of
appearance of density intrusions in the IHS shown in our simulations
repeats periodically in nature. Since the increase of the average density
in the IHS may change the pressure balance between the IHS and OHS, the
effects discussed in our Letter may have consequences for the global structure of the heliosphere.
The density profile in the neighborhood of the HP is an interesting question
in this context as well.
We believe that the transport mechanism revealed by our simulations is
important for a proper understanding of the interaction of the SW
and interstellar plasmas.




\vspace{-0.5cm}
\acknowledgments
M.S. and W.M. acknowledge support by the Polish National Science Center (N N307 0564 40).
R.R. acknowledges support by the Institute of Aviation, HECOLS project, and ISSI.

\clearpage



\clearpage




\begin{thebibliography}{23}
\expandafter\ifx\csname natexlab\endcsname\relax\def\natexlab#1{#1}\fi

\bibitem[{{Balsara} \& {Spicer}(1999)}]{BalSpi99}
{Balsara}, D.~S., \& {Spicer}, D.~S. 1999, J.~Comp. Phys., 149, 270

\bibitem[{{Borovikov} {et~al.}(2008){Borovikov}, {Pogorelov}, {Zank}, \&
  {Kryukov}}]{Boretal08}
{Borovikov}, S.~N., {Pogorelov}, N.~V., {Zank}, G.~P., \& {Kryukov}, I.~A.
  2008, Astrophys. J., 682, 1404

\bibitem[{{Burlaga} \& {Ness}(2012)}]{BurNes12}
{Burlaga}, L.~F., \& {Ness}, N.~F. 2012, Astrophys. J., 749, 13

\bibitem[{Burlaga {et~al.}(2013)Burlaga, Ness, \& Stone}]{Buretal13}
Burlaga, L.~F., Ness, N.~F., \& Stone, E.~C. 2013, Science, 341, 147

\bibitem[{{Czechowski} {et~al.}(2010){Czechowski}, {Strumik}, {Grygorczuk},
  {Grzedzielski}, {Ratkiewicz}, \& {Scherer}}]{Czeetal10b}
{Czechowski}, A., {Strumik}, M., {Grygorczuk}, J., {Grzedzielski}, S.,
  {Ratkiewicz}, R., \& {Scherer}, K. 2010, Astron. Astrophys., 516, A17+

\bibitem[{{Drake} {et~al.}(2010){Drake}, {Opher}, {Swisdak}, \&
  {Chamoun}}]{Draetal10}
{Drake}, J.~F., {Opher}, M., {Swisdak}, M., \& {Chamoun}, J.~N. 2010,
  Astrophys. J., 709, 963

\bibitem[{{Fahr} {et~al.}(1986){Fahr}, {Neutsch}, {Grzedzielski}, {Macek}, \&
  {Ratkiewicz-Landowska}}]{Fahetal86}
{Fahr}, H.~J., {Neutsch}, W., {Grzedzielski}, S., {Macek}, W., \&
  {Ratkiewicz-Landowska}, R. 1986, Space Sci. Rev., 43, 329

\bibitem[{{Florinski} {et~al.}(2005){Florinski}, {Zank}, \&
  {Pogorelov}}]{Floetal05}
{Florinski}, V., {Zank}, G.~P., \& {Pogorelov}, N.~V. 2005, J.~Geophys. Res.,
  110, 7104

\bibitem[{{Gurnett} {et~al.}(2013){Gurnett}, {Kurth}, {Burlaga}, \&
  {Ness}}]{Guretal13}
{Gurnett}, D.~A., {Kurth}, W.~S., {Burlaga}, L.~F., \& {Ness}, N.~F. 2013,
  Science, 341, 1489

\bibitem[{Krimigis {et~al.}(2013)Krimigis, Decker, Roelof, Hill, Armstrong,
  Gloeckler, Hamilton, \& Lanzerotti}]{Krietal13}
Krimigis, S.~M., Decker, R.~B., Roelof, E.~C., Hill, M.~E., Armstrong, T.~P.,
  Gloeckler, G., Hamilton, D.~C., \& Lanzerotti, L.~J. 2013, Science, 341, 144

\bibitem[{{Kurganov} \& {Tadmor}(2000)}]{KurTad00}
{Kurganov}, A., \& {Tadmor}, E. 2000, J.~Comp. Phys., 160, 241

\bibitem[{{Liewer} {et~al.}(1996){Liewer}, {Karmesin}, \&
  {Brackbill}}]{Lieetal96}
{Liewer}, P.~C., {Karmesin}, S.~R., \& {Brackbill}, J.~U. 1996, J.~Geophys.
  Res., 101, 17119

\bibitem[{{Macek} \& {Grzedzielski}(1985)}]{MacGrz85}
{Macek}, W., \& {Grzedzielski}, S. 1985, in Twenty Yaears of Plasma Physics,
  ed. B.~{McNamara}, 320--327

\bibitem[{{Nerney} {et~al.}(1995){Nerney}, {Suess}, \& {Schmahl}}]{Neretal95}
{Nerney}, S., {Suess}, S.~T., \& {Schmahl}, E.~J. 1995, J.~Geophys. Res., 100,
  3463

\bibitem[{{Richardson} \& {Wang}(2012)}]{RicWan12}
{Richardson}, J.~D., \& {Wang}, C. 2012, Astrophys. J. Lett., 759, L19

\bibitem[{{Schoeffler} {et~al.}(2012){Schoeffler}, {Drake}, \&
  {Swisdak}}]{Schetal12}
{Schoeffler}, K.~M., {Drake}, J.~F., \& {Swisdak}, M. 2012, Astrophys. J.
  Lett., 750, L30

\bibitem[{Sonnerup \& Scheible(1998)}]{SonSch98}
Sonnerup, B., \& Scheible, M. 1998, Analysis methods for multi-spacecraft data,
  185

\bibitem[{Stone {et~al.}(2013)Stone, Cummings, McDonald, Heikkila, Lal, \&
  Webber}]{Stoetal13}
Stone, E.~C., Cummings, A.~C., McDonald, F.~B., Heikkila, B.~C., Lal, N., \&
  Webber, W.~R. 2013, Science, 341, 150

\bibitem[{{Strumik} {et~al.}(2013){Strumik}, {Czechowski}, {Grzedzielski},
  {Macek}, \& {Ratkiewicz}}]{Stretal13}
{Strumik}, M., {Czechowski}, A., {Grzedzielski}, S., {Macek}, W.~M., \&
  {Ratkiewicz}, R. 2013, Astrophys. J. Lett., 773, L23

\bibitem[{{Swisdak} {et~al.}(2013){Swisdak}, {Drake}, \& {Opher}}]{Swietal13}
{Swisdak}, M., {Drake}, J.~F., \& {Opher}, M. 2013, Astrophys. J. Lett., 774,
  L8

\bibitem[{{Swisdak} {et~al.}(2010){Swisdak}, {Opher}, {Drake}, \& {Alouani
  Bibi}}]{Swietal10}
{Swisdak}, M., {Opher}, M., {Drake}, J.~F., \& {Alouani Bibi}, F. 2010,
  Astrophys. J., 710, 1769

\bibitem[{{Webber} \& {McDonald}(2013)}]{WebMcD13}
{Webber}, W.~R., \& {McDonald}, F.~B. 2013, Geophys. Res. Lett., 40, 1665

\bibitem[{{Zank} {et~al.}(1996){Zank}, {Pauls}, {Williams}, \&
  {Hall}}]{Zanetal96}
{Zank}, G.~P., {Pauls}, H.~L., {Williams}, L.~L., \& {Hall}, D.~T. 1996,
  J.~Geophys. Res., 101, 21639

\end{thebibliography}
\end{document}